\documentclass[journal]{IEEEtran}
\usepackage{cite,subfigure}
\usepackage{amsmath,amssymb,amsfonts}
\usepackage{algorithmic}
\usepackage{graphicx}
\usepackage{epstopdf}
\usepackage{multirow,array}
\usepackage{url}
\usepackage{dblfloatfix}
\usepackage[dvipsnames]{xcolor}
\usepackage{pdfpages}

\usepackage{todonotes}
\newcommand{\mycomment}[1]{}

\begin{document}

\bstctlcite{IEEE:BSTcontrol}

\title{OEDIPUS: An Experiment Design Framework for Sparsity-Constrained MRI}

\author{Justin P. Haldar,~\IEEEmembership{Senior Member,~IEEE,} and Daeun Kim,~\IEEEmembership{Student Member,~IEEE}
\thanks{J. Haldar and D. Kim are with the Signal and Image Processing Institute, Ming Hsieh Department of Electrical Engineering, University of Southern California, Los Angeles, CA, 90089, USA.}~
\thanks{This work was supported in part by the National Science Foundation (NSF) under CAREER award CCF-1350563 and the National Institutes of Health (NIH) under grants NIH-R21-EB022951, NIH-R01-MH116173, NIH-R01-NS074980, and NIH-R01-NS089212. Computation for some of the work described in this paper was supported by the University of Southern California's Center for High-Performance Computing (http://hpcc.usc.edu/).}}

\maketitle

\begin{abstract}
This paper introduces a new estimation-theoretic framework for experiment design in the context of MR image reconstruction under sparsity constraints.  The new framework is called OEDIPUS (Oracle-based Experiment Design for Imaging Parsimoniously Under Sparsity constraints), and is based on combining the constrained Cram\'{e}r-Rao bound with classical experiment design techniques.  Compared to popular random sampling approaches, OEDIPUS is fully deterministic and automatically tailors the sampling pattern to the specific imaging context of interest (i.e., accounting for coil geometry, anatomy, image contrast, etc.).  OEDIPUS-based experiment designs are evaluated  using retrospectively subsampled \emph{in vivo} MRI data in several different contexts. Results demonstrate that OEDIPUS-based experiment designs have some desirable characteristics relative to  conventional MRI sampling approaches.  
\end{abstract}  


\section{Introduction}
Slow data acquisition speed is a natural consequence of acquiring MRI data with high signal-to-noise ratio (SNR), high spatial resolution, and desirable image contrast characteristics.  Slow acquisitions are problematic because they are expensive, limit scanner throughput and temporal resolution, and can be uncomfortable for scan subjects.   In practice, these concerns lead to the use of shorter-duration experiments that represent undesirable trade-offs between image resolution and contrast when data is sampled according to the conventional sampling theorem \cite{edelstein2010}.  To mitigate this problem, the research community has been working for decades to develop image reconstruction methods that enable conventional sampling requirements to be relaxed  \cite{liang1992}.  While many such image reconstruction methods have been proposed over the years, this paper focuses on a popular and widely-used class of methods based on sparsity constraints \cite{lustig2008}.

While successful applications of sparsity-constrained MRI are widespread throughout the literature, the optimal design of sampling patterns remains a longstanding open problem.  Because there is already a large amount of  literature on this topic, it would be impractical for us to comprehensively review every contribution.  Instead,  our literature review focuses only on what we believe is the most relevant previous work.

Some of the earliest sparsity-based methods sample low-resolution data at the conventional Nyquist rate, and use sparsity constraints to recover the unsampled high-frequency data and enable super-resolution reconstruction \cite{liang1989}.  More recently, insight provided by compressed sensing theory has inspired the emergence and popularity of pseudo-random Fourier sampling schemes \cite{lustig2007}.  

Random Fourier sampling for MRI has often been motivated by theoretical performance guarantees for random sampling that depend on mathematical concepts of incoherence and/or restricted isometry \cite{donoho2006,candes2008b}. Within the MRI community, a fairly common misconception is that random sampling is required for good sparsity-constrained MRI.  In reality, random Fourier sampling is neither necessary nor sufficient for good performance \cite{haldar2010a,adcock2017,lustig2007}, particularly in the case of multi-channel imaging for which the reconstruction problem is often not actually underdetermined (and is instead just ill-posed). Most modern approaches have come to rely on heuristic variations of random Fourier sampling that are often tuned for each new application context based on extensive empirical trial-and-error testing \cite{lustig2007,wang2010,knoll2011a,zijlstra2016}.  While these kinds of random sampling variations often work adequately well, their good performance is not guaranteed from a theoretical perspective \cite{haldar2010a}, empirical tuning of random sampling characteristics is often onerous, and it remains unclear whether the results are near-optimal.

This paper describes a principled method called OEDIPUS (Oracle-based Experiment Design for Imaging Parsimoniously Under Sparsity constraints) for designing sampling patterns for sparsity-constrained MRI.  Compared to randomization-based approaches, OEDIPUS is fully deterministic and automatically tailors the sampling pattern to the specific imaging context (accounting for coil geometry, anatomy, image contrast, etc.) of interest without the need for manual interaction.  A preliminary description of this work was previously presented in \cite{haldar2017}. 

The OEDIPUS framework is inspired by classical experiment design approaches. Experiment design is a well-developed subfield of statistics \cite{federov1972,silvey1980,pukelsheim1993}, and experiment design tools have been previously used to improve the efficiency of a wide variety of MRI experiments. These methods are perhaps most visible within the MRI literature in the context of selecting flip angles, repetition times, echo times, and related pulse sequence parameters that influence image contrast. Such methods have already had a major impact in the context of quantitative MRI parameter estimation for applications such as spectroscopy \cite{cavassila2001}, water-fat separation \cite{pineda2005}, dynamic contrast enhanced imaging \cite{denaeyer2011}, diffusion MRI \cite{alexander2008}, relaxometry  \cite{jones1996}, and multi-parametric mapping \cite{zhao2017}, among many others.  

However, the use of experiment design techniques to optimize k-space sampling patterns (as done in OEDIPUS) has not been as widely explored, in part due to the computational complexity of applying experiment design techniques to image reconstruction problems (which are often substantially larger in scale than MRI parameter estimation problems).  There are a few notable exceptions.  Marseille \emph{et al.} \cite{marseille1994} performed optimization of phase encoding positions in 1D  under a heuristic  surrogate model of an MR image.  Reeves \emph{et al.} \cite{reeves1995,gao2000,gao2000a,gao2001,gao2006} optimized phase encoding positions in 2D assuming that the image had limited support in the spatial domain, with perfect \emph{a priori} knowledge of the image support, while later authors explored a similar approach in the context of dynamic imaging \cite{sharif2010}.  Xu \emph{et al.} \cite{xu2005} performed optimization of phase encoding positions in 1D for the case of parallel imaging using the SENSE model \cite{pruessmann1999}. Samsonov \cite{samsonov2009} optimized the projection angles for  radial imaging under the SENSE model.  Haldar \emph{ et al.} \cite{haldar2009a} compared different spatiotemporal (k-t) sampling patterns for parameter mapping in the presence of prior knowledge of the image contrast model. Levine and Hargreaves \cite{levine2018} optimized 2D sampling patterns under support constraints and parallel imaging constraints with the SENSE model.  Importantly, none of these approaches directly incorporated transform domain sparsity constraints into the experiment design problem.

The only previous sparsity-based MR experiment design work we are aware of is by Seeger \emph{et al.} \cite{seeger2010}.  While this work has the strongest similarity to OEDIPUS among existing methods from the literature, it also has substantial differences, including being developed with a completely different context in mind.  In particular, the approach described by Seeger \emph{et al.} is intended for real-time adaptive  sampling design, in which  each subsequent k-space sampling location is chosen greedily on-the-fly using real-time feedback from the data that has already been acquired for that scan.  In contrast, OEDIPUS is a non-adaptive and subject-independent approach to experiment design, which enables offline prior computation of the sampling pattern and reuse of the same sampling scheme for different subjects.  Another major difference is that Seeger \emph{et al.} assume a Bayesian formulation in which the sparse image coefficients are modeled using the Laplace distribution as a statistical prior, while OEDIPUS is a non-Bayesian approach that does not use any statistical priors.  Instead, OEDIPUS takes a deterministic approach based on the constrained Cram\'{e}r-Rao bound (CRB) \cite{gorman1990,ben-haim2010a}.

The rest of this paper is organized as follows.  Section~\ref{sec:crb} introduces notation, reviews the CRB for linear inverse problems with Gaussian noise, and reviews optimal experiment design approaches that aim to minimize the CRB.  Section~\ref{sec:oedipus} describes the  general OEDIPUS framework as an extension of the methods from Section~\ref{sec:crb}.  This section also describes the specific OEDIPUS implementation choices we have used in the remainder of the paper.  Section~\ref{sec:results} describes empirical evaluations of OEDIPUS with comparisons against other common k-space sampling methods from the literature.  Discussion and  conclusions are presented in Section~\ref{sec:disc}.

\section{Review of ``Classical" Experiment Design}
\label{sec:crb}

\subsection{Linear Inverse Problems with Gaussian Noise}
The data acquisition process in MRI is often formulated using a standard finite-dimensional linear model \cite{fessler2010b}
\begin{equation}
\mathbf{d} = \mathbf{A} \mathbf{f} + \mathbf{n},\label{eq:lin}
\end{equation}
where $\mathbf{f}\in \mathbb{C}^N$ are the coefficients of a finite-dimensional representation of the image, $\mathbf{d} \in \mathbb{C}^M$ is the vector of measured data samples, the system matrix $\mathbf{A} \in \mathbb{C}^{M \times N}$ provides a linear model of the MRI data acquisition physics (e.g., potentially including Fourier encoding, sensitivity encoding, and physical effects like field inhomogeneity), and $\mathbf{n} \in \mathbb{C}^M$ is modeled as zero-mean circularly-symmetric complex Gaussian noise.  Without loss of generality, we will assume that noise whitening has been applied such that the entries of $\mathbf{n}$ are assumed to be independent and identically distributed with variance $\sigma^2=1$.  

The objective of MRI reconstruction methods is to estimate the original image $\mathbf{f}$ based only on the noisy measured data $\mathbf{d}$ and prior knowledge of the noise statistics and imaging model $\mathbf{A}$.  We will denote such an estimator as $\hat{\mathbf{f}}(\mathbf{d})$.

\subsection{The Cram\'{e}r-Rao Bound for the Linear Model}
The CRB is a theoretical lower bound on the covariance of an unbiased estimator \cite{kay1993}.  Assume for now that data has been acquired according to Eq.~\eqref{eq:lin}, that $\mathbf{A}$ has full-column rank, and that we have an unbiased estimator $\hat{\mathbf{f}}(\mathbf{d})$, i.e., $E[\hat{\mathbf{f}}(\mathbf{d})] = \mathbf{f}$, where $E[\cdot]$ denotes statistical expectation.  The CRB  says that
\begin{equation}
\mathrm{cov}(\hat{\mathbf{f}}(\mathbf{d})) \succeq  (\mathbf{A}^H\mathbf{A})^{-1},\label{eq:loew}
\end{equation}
where $\mathrm{cov}(\hat{\mathbf{f}}(\mathbf{d})) \in \mathbb{C}^{N\times N}$ is the covariance matrix of the estimator $\hat{\mathbf{f}}(\mathbf{d})$, and the $\succeq$ symbol is used to signify an inequality relationship with respect to the Loewner ordering of positive semidefinite matrices \cite{pukelsheim1993}.  To be precise, Eq.~\eqref{eq:loew} signifies that $\mathrm{cov}(\hat{\mathbf{f}}(\mathbf{d}))- (\mathbf{A}^H\mathbf{A})^{-1}$ will always be a positive semidefinite matrix.  As a result, the CRB matrix $(\mathbf{A}^H\mathbf{A})^{-1}$ represents the smallest possible covariance matrix that can be achieved using an unbiased estimator, since we must always have that $\mathbf{u}^H\mathrm{cov}(\hat{\mathbf{f}}(\mathbf{d}))\mathbf{u} \geq  \mathbf{u}^H(\mathbf{A}^H\mathbf{A})^{-1}\mathbf{u}$ for every possible choice of $\mathbf{u} \in \mathbb{C}^N$.  To illustrate the importance of this relationship, consider the case where we select $\mathbf{u}$ to be the $j$th canonical basis vector for $\mathbb{C}^N$, i.e., the $j$th entry of $\mathbf{u}$ is $[\mathbf{u}]_j = 1$, with $[\mathbf{u}]_n = 0$ for all $n\neq j$.  This implies that $\mathrm{var}\left([\hat{\mathbf{f}}(\mathbf{d})]_j\right) \geq  \left[ (\mathbf{A}^H\mathbf{A})^{-1} \right]_{jj}$, where $\mathrm{var}([\hat{\mathbf{f}}(\mathbf{d})]_j)$ is the scalar variance of the estimator for the $j$th entry of $\mathbf{f}$.  Note that, because we have chosen $j$ arbitrarily in this illustration, we have that the $N$ diagonal elements of the CRB matrix each provide a fundamental lower bound on the variance for the corresponding $N$ entries of the unbiased estimator $\hat{\mathbf{f}}(\mathbf{d})$.  

Importantly, the CRB is a function of the matrix $\mathbf{A}$, which means that it can be possible to use the CRB to compare the SNR-efficiency of different experiments.  For readers familiar with the parallel MRI literature, it is worth mentioning that in the case of SENSE reconstruction, the CRB is an unnormalized version of the SENSE g-factor \cite{pruessmann1999,robson2008}, which is already well known and widely used for its usefulness in comparing different k-space sampling strategies in parallel MRI.  

In addition, the CRB is a reconstruction-agnostic lower bound, in the sense that it is completely independent from the actual method used to estimate $\mathbf{f}$.  However, it should also be noted that the CRB is only valid for unbiased estimators, while different estimation-theoretic tools are necessary to characterize biased estimators.  This fact is particularly relevant to this paper, since most sparsity-constrained reconstruction methods use some form of regularization procedure, and regularized image reconstruction will generally produce biased results \cite{fessler1996a}.  Nevertheless, minimizing the CRB can still be a useful approach to experiment design even if a biased estimator will  ultimately be used for reconstruction.  In some sense, minimizing the CRB ensures that the data acquisition procedure represented by the matrix $\mathbf{A}$ will encode as much information as possible about $\mathbf{f}$.

\subsection{Optimal Experiment Design using the CRB}
Methods for minimizing the CRB have been studied for many years in both statistics \cite{federov1972,silvey1980,pukelsheim1993} and  signal processing \cite{reeves1999,broughton2010}.  There are generally two classes of popular experiment design approaches: continuous approaches and discrete approaches.  Optimal continuous designs are often easier to compute than optimal discrete designs, but are approximate in nature and do not translate easily to the constraints of practical MRI experiments because they do not account for the fact that real experiments acquire a finite, integer number of samples.  As a result, we will focus on discrete designs in this paper.

The discrete version of  optimal experiment design is often formulated as follows:  given a set of $P$ candidate measurements (with $P > M$), choose a subset of $M$ measurements in a way that minimizes the CRB.  Formally, let $\mathbf{a}_p \in \mathbb{C}^{1\times N}$ for $p=1,\ldots,P$ represent the $P$ potential rows of the optimal $\mathbf{A}$ matrix from Eq.~\eqref{eq:lin}.   Given the corresponding set of potential measurement indices $\Gamma = \{1, \ldots, P\}$, the optimal design problem is to select a subset $\Omega \subset \Gamma $ of cardinality $M$ to ensure that the resulting CRB is as small as possible.  This leads to the optimal design problem
\begin{equation}
\Omega^* = \arg\min_{\substack{\Omega \subset \Gamma \\ |\Omega| = M}} J\left( \left(\sum_{m \in \Omega} \mathbf{a}_m^H \mathbf{a}_m \right)^{-1}\right), \label{eq:cost}
\end{equation}
where $J(\cdot): \mathbb{C}^{N\times N} \rightarrow \mathbb{R}$ is a user-selected functional used to measure the size of the CRB.  

An ideal choice of $J(\cdot)$ would ensure that for arbitrary positive semidefinite matrices $\mathbf{C} \in \mathbb{C}^{N\times N}$ and $\mathbf{D} \in \mathbb{C}^{N\times N}$, having $J(\mathbf{C}) \geq J(\mathbf{D})$ would imply that $\mathbf{C} \succeq \mathbf{D}$, and therefore that $\mathbf{D}$ is uniformly smaller than $\mathbf{C}$.  Unfortunately, the Loewner ordering is only a partial ordering, and not all positive semidefinite matrices are comparable.  This means that there are generally no universally-optimal experiment designs, and it is necessary to choose $J(\cdot)$ subjectively \cite{pukelsheim1993}.  

Out of many options \cite{pukelsheim1993}, one of the most popular choices is the \emph{average variance} criterion (also known as \emph{A-optimality}):
\begin{equation}
\begin{split}
J\left( \left(\sum_{m \in \Omega} \mathbf{a}_m^H \mathbf{a}_m \right)^{-1}\right) \triangleq\mathrm{Trace}\left[\left(\sum_{m \in \Omega} \mathbf{a}_m^H \mathbf{a}_m \right)^{-1}\right],
\end{split}
\end{equation}
which takes its name from the fact that the cost function is a lower bound on the sum of elementwise variances, i.e.,  
\begin{equation}
\mathrm{Trace}\left[\left(\sum_{m \in \Omega} \mathbf{a}_m^H \mathbf{a}_m \right)^{-1}\right] \leq \sum_{n=1}^N \mathrm{var}([\hat{\mathbf{f}}(\mathbf{d})]_n)\label{eq:mse}
\end{equation}
for every possible unbiased estimator $\hat{\mathbf{f}}(\mathbf{d})$.  Since the mean-squared error (MSE) is proportional to the sum of elementwise variances for any unbiased estimator, Eq.~\eqref{eq:mse} also provides a CRB-based lower bound for the MSE.  For the sake of concreteness and without loss of generality, we will focus on the A-optimality criterion in this paper, though note that OEDIPUS can also easily be used with some of the other cost functionals that are popular in the experiment design literature.

Given this choice of  $J(\cdot)$, the optimization problem from Eq.~\eqref{eq:cost} is still nontrivial to solve because it is a nonconvex integer programming problem, and global optimality may require exhaustive search through the set of candidate solutions.  This is generally infeasible from a practical computational perspective.  Specifically, there are $P$-choose-$M$ possible designs $\Omega$, and this will be a very large number in most scenarios of interest.  As a result, Eq.~\eqref{eq:cost} is often minimized using heuristic algorithms that can provide high-quality solutions, but which cannot guarantee global optimality.  Popular heuristics include the use of Monte Carlo methods (which randomly construct potential sampling pattern candidates for evaluation) or greedy algorithms (which add, subtract, or exchange candidate measurements to/from the current design in a sequential greedy fashion) \cite{federov1972,silvey1980,reeves1999,broughton2010}.  

While these kinds of algorithms have been used previously for k-space trajectory design in several image reconstruction contexts \cite{reeves1995,gao2000,gao2000a,gao2001,gao2006,xu2005}, certain computational problems arise when the CRB is applied to conventional image reconstruction approaches (without sparsity constraints).  In particular, the CRB matrix is of size $N\times N$, which can be very large if $N$ is chosen equal to the number of voxels in a reasonably-sized image.  For example, for a $256 \times 256$ image, the CRB matrix will have nearly $4.3\times 10^9$ entries, which will occupy more than 34 GB of memory if each entry is stored in a standard double precision floating point number format. While storing a matrix of this size in memory is feasible on modern high-end computers, other computations (e.g., inverting the matrix) are generally prohibitively expensive. Another problem is that the assumptions underlying the CRB will be violated whenever $N > M$, which means that traditional experiment design approaches are not straightforward to apply for the design of undersampled experiments.  These issues are mitigated within the OEDIPUS framework as described in the sequel.

\section{Oracle-based Experiment Design for Imaging Parsimoniously Under Sparsity constraints}\label{sec:oedipus}

\subsection{Theoretical components of OEDIPUS}\label{sec:theory}
The previous section considered CRB-based design under the standard unconstrained linear model, but modifications are necessary to inform experiment design under sparsity constraints.  To describe OEDIPUS, we will assume that it is known in advance that the image vector $\mathbf{f}$ possesses a certain level of sparsity in a known transform domain, i.e., $\|\boldsymbol{\Psi}\mathbf{f}\|_0 \leq S$  where $\boldsymbol{\Psi} \in \mathbb{C}^{Q \times N}$ is a known sparsifying transform, $S$ is the known sparsity level, and the $\ell_0$-``norm" $\|\cdot\|_0$ counts the number of nonzero entries of its vector argument.  It should be noted that the assumptions that the signal is strictly sparse and that the value of $S$ is known in advance are not conventional in sparsity-constrained MRI reconstruction, though we need them to construct the CRB for this case.  

In this work, we will make the additional assumption that $\boldsymbol{\Psi}$ is left-invertible such that $\mathbf{f} = \boldsymbol{\Psi}^\dagger\boldsymbol{\Psi}\mathbf{f}$ for all $\mathbf{f} \in \mathbb{C}^N$, where $^\dagger$ is used to denote the matrix pseudoinverse.  In the presence of sparsity constraints, this allows us to rewrite Eq.~\eqref{eq:lin} in terms of the vector of transform coefficients $\mathbf{c} = \boldsymbol{\Psi}\mathbf{f} \in \mathbb{C}^Q$  as:
\begin{equation}
\mathbf{d} = \mathbf{A}\boldsymbol{\Psi}^\dagger\mathbf{c} + \mathbf{n}.\label{eq:sparse}
\end{equation}
Recently, Ben-Haim and Eldar \cite{ben-haim2010a} derived a version of the constrained CRB that is applicable to Eq.~\eqref{eq:sparse} under the constraint that $\|\mathbf{c}\|_0 = S$.  In particular, they derived that the CRB in this case is equal to the CRB of the \emph{oracle estimator}, i.e., the estimator that is additionally bestowed  with perfect knowledge of which of the $S$ entries of $\mathbf{c}$ are non-zero.  

The oracle-based version of the problem can be viewed as a simple unconstrained linear model, and thus fits directly within the framework described in the previous section.  In particular, let $\tilde{\mathbf{c}} \in \mathbb{C}^S$ denote the vector of nonzero entries from the $\mathbf{c}$ vector, with the two vectors related through $\mathbf{c} = \mathbf{U}\tilde{\mathbf{c}}$, where $\mathbf{U} \in \mathbb{R}^{Q \times S}$ is a matrix formed by concatenating the $S$ columns of the $Q\times Q$ identity matrix corresponding to the non-zero entries of $\mathbf{c}$.  This matrix  can be viewed as performing a simple zero-filling operation  (i.e., putting the nonzero entries of $\tilde{\mathbf{c}}$ in their correct corresponding positions and leaving the remaining entries equal to zero).  Given this notation, the oracle-based version of the problem is then written as
\begin{equation}
\mathbf{d} = \mathbf{A}\boldsymbol{\Psi}^\dagger \mathbf{U} \tilde{\mathbf{c}} + \mathbf{n}.
\end{equation}
From here, we can use the results of the previous section to easily derive the oracle-based CRB for $\hat{\tilde{\mathbf{c}}}(\mathbf{d})$:
\begin{equation}
\mathrm{cov}(\hat{\tilde{\mathbf{c}}}(\mathbf{d})) \succeq \left(\mathbf{U}^H\left(\boldsymbol{\Psi}^\dagger\right)^H \mathbf{A}^H\mathbf{A}\boldsymbol{\Psi}^\dagger \mathbf{U} \right)^{-1},\label{eq:ctilde}
\end{equation}
such that the oracle-based CRB for $\hat{\mathbf{c}}(\mathbf{d}) = \mathbf{U}\hat{\tilde{\mathbf{c}}}(\mathbf{d})$ is:
\begin{equation}
\mathrm{cov}(\hat{\mathbf{c}}(\mathbf{d})) \succeq \mathbf{U}\left(\mathbf{U}^H\left(\boldsymbol{\Psi}^\dagger\right)^H \mathbf{A}^H\mathbf{A}\boldsymbol{\Psi}^\dagger \mathbf{U} \right)^{-1}\mathbf{U}^H,\label{eq:c}
\end{equation}
and the oracle-based CRB for $\hat{f}(\mathbf{d}) = \boldsymbol{\Psi}^\dagger\hat{\mathbf{c}}(\mathbf{d})$ is:
\begin{equation}
\begin{split}
\mathrm{cov}&(\hat{\mathbf{f}}(\mathbf{d})) \succeq \\ &\boldsymbol{\Psi}^\dagger\mathbf{U}\left(\mathbf{U}^H\left(\boldsymbol{\Psi}^\dagger\right)^H \mathbf{A}^H\mathbf{A}\boldsymbol{\Psi}^\dagger \mathbf{U} \right)^{-1}\mathbf{U}^H\left(\boldsymbol{\Psi}^\dagger\right)^H.
\end{split}\label{eq:f}
\end{equation}
Given these sparsity-constrained CRBs, it is easy to formulate and solve an experiment design problem using exactly the same design methods described for the linear model in the previous section.  The CRB expressions given above will be valid whenever the $M \times S$ matrix $\mathbf{A} \boldsymbol{\Psi}^\dagger\mathbf{U}$ has full column rank, which  is still feasible for undersampled experiments where $M < N$ (though we still need to satisfy $S \leq M$).  

In addition to enabling the use of sparsity constraints and undersampling in experiment design, the expressions in Eqs.~\eqref{eq:ctilde}-\eqref{eq:f} are generally also substantially easier to work with from a computational point of view.  In particular, it should be noted that the matrices being inverted in these expressions are only of size $S\times S$, rather than the $N \times N$ matrices that needed to be inverted in the previous case.  This can lead to massive computational complexity advantages since we usually have $ S \ll N$.  For example, for a 256$\times$256 image with $S = 0.15 N$, the matrix to be inverted in this case would have roughly 2.25\% the memory footprint of the matrix to be inverted in the $N\times N$ case.  In this example, this is the difference between a 0.8 GB matrix and a 34 GB matrix.

Even though the final CRB matrices in Eqs.~\eqref{eq:c} and \eqref{eq:f} are generally larger (respectively of size $Q\times Q$ and $N\times N$) than the $S\times S$ matrix being inverted in Eq.~\eqref{eq:ctilde}, it is not hard to show that the matrices in Eqs.~\eqref{eq:ctilde} and \eqref{eq:c} always have the same trace value.  In addition, the trace value for the matrix from Eq.~\eqref{eq:f} will also have this same trace value if $\boldsymbol{\Psi}$ is a unitary transform (e.g., an orthonormal wavelet transform) and will be proportional to this same trace value if $\boldsymbol{\Psi}$ corresponds to a tight frame representation \cite{kovacevic2007} (e.g., the undecimated Haar wavelet transform \cite{starck2007} or the discrete curvelet transform \cite{candes2006c }).  As a result, it often suffices to work directly with the small $S\times S$ matrix from Eq.~\eqref{eq:ctilde} when performing experiment design using the A-optimality design criterion.

\subsection{Practical Implementation for MRI Applications}
\subsubsection{Constructing the OEDIPUS optimization problem}

For the sake of concreteness and without loss of generality, we will describe our implementation of OEDIPUS assuming that we are applying the A-optimality design criterion to the sparsity-constrained CRB from Eq.~\eqref{eq:ctilde}.  Given an application context, there are several ingredients we need to select prior to the experiment design process.

The first choice that needs to be made is the sparsifying transform $\boldsymbol{\Psi}$.  While some groups have explored the use of data-dependent sparsifying transforms (e.g., \cite{ravishankar2011,lingala2013}), it is by far more common for the sparsifying transform to be chosen in advance based on past empirical experience with a given MRI application.  Our description will focus on the conventional case where $\boldsymbol{\Psi}$ is fixed and provided in advance.

The second choice we need to make relates to the set of candidate measurement vectors $\mathbf{a}_p \in \mathbb{C}^{1\times N}$ for $p=1,\ldots,P$.  This choice is relatively straightforward in the case of ideal single-channel Fourier encoding.  Given a field-of-view (FOV) size and the number of voxels within the FOV (which we assume are fixed, and dictated by the specific application context of interest), the candidate measurement vectors will often have entries (assuming an $N$-dimensional voxel-based finite series expansion of the image \cite{fessler2010b}) of the form:
\begin{equation}
[\mathbf{a}_p]_{1n} = b_{pn} \exp(-i2\pi \mathbf{k}_p\cdot \mathbf{r}_n),
\end{equation}
where $b_{pn}$ is a weighting coefficient that depends on the specific choice of voxel basis function (common choices are to use either a Dirac delta voxel function in which $b_{pn}=1$ or a rectangular voxel basis function in which case $b_{pn}$ takes the form of a sinc function in k-space \cite{fessler2010b}), $\mathbf{r}_n$ is the spatial coordinate vector specifying the center position of the $n$th voxel (the common choice is to choose the voxel centers to lie on a rectilinear grid), and $\mathbf{k}_p$ is the $p$th candidate k-space measurement position.  The parameters $b_{pn}$ and $\mathbf{r}_n$ are parameters of the image model that are generally chosen in advance, and which will be invariant to different experiment design choices.   As a result, selection of the full set of the candidate measurements reduces to selection of the set of $\mathbf{k}_p$ k-space sampling locations.  The set of candidate k-space sampling locations will be sequence dependent, but typical choices might include the set of all possible k-space sampling locations from a ``fully-sampled" Cartesian grid, a ``fully-sampled" radial trajectory, a ``fully-sampled" spiral trajectory, etc.  In each of these cases, we can define ``full-sampling" either based on the conventional Nyquist rate, or we can also consider schemes that sample more densely than the Nyquist rate if we want to consider a larger range of possibilities.  There is a trade-off here --  the more options we consider, the better our optimized schemes can be, though this comes at the expense of increased computational complexity.  

Some additional considerations are required if the model is designed to incorporate sensitivity encoding, field inhomogeneity, radiofrequency encoding, or other similar effects.  For illustration, in the case of parallel imaging with the SENSE model \cite{pruessmann1999}, the measurements take the more general form:
\begin{equation}
[\mathbf{a}_p]_{1n} = c_p(\mathbf{r}_n)b_{pn} \exp(-i2\pi \mathbf{k}_p\cdot \mathbf{r}_n),
\end{equation}
where $c_p(\mathbf{r}_n)$  is the value of the sensitivity profile at spatial location $\mathbf{r}_n$ for the coil used in the $p$th measurement.\footnote{In parallel MRI, the same k-space location is measured simultaneously through all of the coils that are present in the receiver array.  We have adopted a notation in which each individual measurement is indexed by a separate $p$ value, even if some measurements are required to be sampled simultaneously.}\mycomment{AE.4} The main complication in applying OEDIPUS to the SENSE model is that, while the Fourier encoding model is independent of the person or object being  imaged, the coil sensitivity profiles often vary slightly from subject to subject, and this means that ideally we should design optimal experiments on a subject-by-subject basis.  However, subject-by-subject design is not conducive to the kind of offline subject-independent experiment design that we would like to use OEDIPUS to address.  Similar complications exist because of  subject-dependent effects for other advanced modeling schemes that account for $B_0$ inhomogeneity, transmit $B_1$ inohomogeneity, etc.  To mitigate the issue of subject-dependent observations, our formulation for the SENSE case (and similarly for other subject-dependent acquisition models) will assume that we have a set of $T$ different representative sensitivity maps that are appropriate for the given application context, i.e.,
\begin{equation}
[\mathbf{a}_p^t]_{1n} = c_p^t(\mathbf{r}_n)b_{pn} \exp(-i2\pi \mathbf{k}_p\cdot \mathbf{r}_n),
\end{equation}
for $t = 1, \ldots, T$,  and will optimize the experiment design with respect to the ensemble of these representative cases.  We will formalize this approach mathematically later in this section.

The remaining ingredients that need to be chosen are the sparsity level $S$ and the matrix $\mathbf{U}$ that specifies the transform-domain positions of the non-zero coefficients.  However, real MRI images are only approximately sparse rather than exactly sparse, and the locations of the significant transform-domain coefficients will generally vary from subject to subject.  To address these issues and enable subject-independent experiment design, we will assume that we are given a set of $K$ exemplar images acquired from a given imaging context, and similar to the previous case, will identify $S_k$ and $\mathbf{U}_k$ values for $k=1,\ldots,K$ that are appropriate for each case.  Ultimately, we will attempt to find a design that is optimal with respect to the ensemble of exemplars.  In this context, we first apply the sparsifying transform $\boldsymbol{\Psi}$ to each exemplar, and explore sparse approximation quality by hard-thresholding the transform domain coefficients with different choices of $S_k$.  The value of $S_k$ is then chosen as the smallest value  that gives adequate sparse approximation quality for that exemplar.  Note that ``adequate quality" is necessarily subjective and application dependent.  Given the choice of $S_k$ for each exemplar image, we can obtain corresponding $\mathbf{U}_k \in \mathbb{R}^{Q\times S_k}$ matrices by identifying the locations of the $S_k$ largest transform coefficients.

We are now almost ready to perform experiment design, except that we have not specified how to handle optimization with respect to the ensemble of  $T$ different measurement models and $K$ different exemplar sparsity constraints.  There are a few different options for obtaining a unified treatment.  One typical approach would be to minimize the average-case CRB across each of these different conditions, which results in the following optimization problem:
\begin{equation}
\Omega^* = \arg\min_{\substack{\Omega \subset \Gamma \\ |\Omega| = M}} \sum_{k=1}^K\sum_{t=1}^T \mathrm{Trace}\left[ \mathbf{C}_{kt}(\Omega) \right], \label{eq:cost2}
\end{equation}
where 
\begin{equation}
\mathbf{C}_{kt}(\Omega) \triangleq  \left(\sum_{m \in \Omega}  \mathbf{U}_k^H \left(\boldsymbol{\Psi}^\dagger\right)^H (\mathbf{a}_m^t)^H \mathbf{a}_m^t \boldsymbol{\Psi}^\dagger \mathbf{U}_k \right)^{-1}.
\end{equation}
Another typical choice would be to optimize the worst-case CRB, which leads to the following minimax optimization problem:
\begin{equation}
\Omega^* = \arg\min_{\substack{\Omega \subset \Gamma \\ |\Omega| = M}}  \max_{(k,t)} \mathrm{Trace}\left[ \mathbf{C}_{kt}(\Omega) \right]. \label{eq:cost3}
\end{equation}
Many other choices are also possible and may be preferable in certain contexts, though we leave such an exploration for future work.  The choice between minimizing the average-case and the worst-case in experiment design is subjective in nature, and will depend on context-dependent factors.  For example, in a large population study with a neuroscientific objective,  experimenters may be able to tolerate the situation where a few subjects have low-quality images if the average image quality for the remaining subjects is very high on average.  On the other hand, in an emergency medical situation, it may be preferable that every individual image meets certain quality standards, even if that means that the average-case image quality is lower as a result.  In the case where $K=T=1$, both Eq.~\eqref{eq:cost2} and \eqref{eq:cost3} lead to identical optimization problems.  

\subsubsection{Solving the OEDIPUS optimization problem}

Given a choice of objective function, it remains to find an experiment design that is as close to optimal as possible.  While there are many potential ways to minimize these cost functionals, the implementation we have used for this paper is based on a modified sequential backward-selection (SBS) procedure.  The standard SBS procedure \cite{reeves1999, broughton2010} starts with a hypothetical experiment that collects all $P$ candidate measurements, and then identifies the individual measurement that is the least important to the CRB and deletes it in a greedy fashion.  This procedure is then iterated, deleting  candidate measurements one-by-one until only $M$ measurement vectors remain.  SBS is greedy and will not generally find the globally optimal solution, but it is a simple deterministic algorithm that typically obtains local minima that compare very well in ultimate quality compared to alternative optimization approaches.  Additional theoretical comments about the characteristics of SBS can be found in \cite{reeves1999}, and methods that have the potential to improve on SBS results are discussed in \cite{broughton2010}.

While the conventional SBS approach may be well-suited to certain applications, it is not immediately applicable to  many MRI contexts in which sets of measurements must be acquired simultaneously.  For example, in parallel imaging, it does not make any sense to delete a k-space measurement made with one coil without also deleting the same k-space measurement from the remaining coils.  Similarly, it would not make sense to delete one k-space measurement from a given readout without also deleting the other measurements from the  same readout.  As a result, it is important to be able to adapt SBS to consider sets of candidate measurements rather than individual candidate measurements.   Specifically (and similar to \cite{gao2001,xu2005,gao2006}), we will assume that our set of $P$ measurements is grouped into $L$ disjoint subsets $\Xi_\ell$ for $\ell = 1,\ldots,L$, and that all measurements within a given subset $\Xi_\ell$ must either be acquired simultaneously or not acquired at all.  We will further assume the practical scenario in which all of these measurement sets have the same cardinality, i.e., $|\Xi_\ell| = C$ for some integer $C$.  This leads to the following modified SBS procedure for OEDIPUS:
\begin{enumerate}
\item  For $k=1,\ldots,K$ and $t=1,\ldots,T$, construct the sparsity-constrained CRB matrix corresponding to the use of all $P$ candidate measurements:
\begin{equation}
\mathbf{C}_{kt}^{\mathrm{full}} =  \left(\sum_{p =1}^P  \mathbf{U}_k^H \left(\boldsymbol{\Psi}^\dagger\right)^H (\mathbf{a}_p^t)^H \mathbf{a}_p^t \boldsymbol{\Psi}^\dagger \mathbf{U}_k \right)^{-1}.\label{eq:full}
\end{equation}
As noted in Sec.~\ref{sec:theory}, this matrix is often not very large (i.e., of size $S\times S$), and will therefore fit easily within the memory capacities of most modern computers if $S$ is chosen suitably small.
\item For each $\ell=1,\ldots,L$, compute the cost-function value (e.g., the average-case cost from Eq.~\eqref{eq:cost2} or the worst-case cost from Eq.~\eqref{eq:cost3}) for an experiment  that uses all $P$ measurements except the $C$ candidate measurements included in $\Xi_\ell$.  
\item The cost-function values are compared for each choice of $\ell$, and the measurement set whose removal causes the smallest degradation in the cost function value is then removed from the set of measurement candidates. Specifically, if $i$ is the index for the removed set of measurement candidates, set $\Gamma = \Gamma \setminus \Xi_i$, and correspondingly set $P = P-C$.  Similarly, remove $\Xi_i$ from the set of measurement candidates such that $L = L-1$.  
\item Repeat Steps 1-3 until the desired number of measurements is achieved, i.e., stop when $|\Gamma| = M$.
\end {enumerate}

The main computational complexity associated with this modified SBS approach is that Step 2 requires $TLK$ matrix inversions at each iteration.   However, as thoroughly discussed by \cite{reeves1999}, it should be noted that it is easy and computationally-efficient to use the Sherman-Morrison-Woodbury matrix inversion lemma \cite{golub1996} to use the matrix from Step 1 to quickly compute the matrices needed for Step 2.  Specifically, if we use $\mathbf{B}_{\ell t} \in \mathbb{C}^{C \times N}$ to denote the matrix formed by stacking together the potential rows associated with $\Xi_\ell$ for the $t$th acquisition model and let $\tilde{\mathbf{B}}_{\ell t}^k = \mathbf{B}_{\ell t} \boldsymbol{\Psi}^\dagger\mathbf{U}_k \in \mathbb{C}^{C \times S}$, then the CRB matrix $\mathbf{C}_{kt}^{\mathrm{full} \setminus \Xi_\ell}$ that would be obtained by keeping all $P$ measurements except those from $\Xi_\ell$ can be written as
\begin{equation}
\begin{split}
\mathbf{C}_{kt}^{\mathrm{full} \setminus \Xi_\ell}
&  =\left(\sum^P_{\substack{p =1 \\ p \notin \Xi_\ell}}  \mathbf{U}_k^H \left(\boldsymbol{\Psi}^\dagger\right)^H (\mathbf{a}_p^t)^H \mathbf{a}_p^t \boldsymbol{\Psi}^\dagger \mathbf{U}_k \right)^{-1}\\
 &= \mathbf{C}_{kt}^{\mathrm{full}} +  \mathbf{G}_{\ell t}^k\left(\mathbf{I}_C - \tilde{\mathbf{B}}^k_{\ell t} \mathbf{G}_{\ell t}^k\right)^{-1} \left(\mathbf{G}_{\ell t}^k\right)^H, 
 \end{split}\label{eq:smw}
\end{equation}
where $\mathbf{I}_C$ is the $C \times C$ identity matrix and 
$\mathbf{G}_{\ell t}^k = \mathbf{C}_{kt}^{\mathrm{full}}  \left(\tilde{\mathbf{B}}_{\ell t}^k\right)^H$.  Note that this final expression only requires the inversion of a $C \times C$ matrix, which is a much faster computation than inverting an $S\times S$ matrix if $C \ll S$ (which it usually will be).  In addition, it should be noted that except for the first time that $\mathbf{C}_{kt}^{\mathrm{full}}$ is computed using Eq.~\eqref{eq:full} in Step 1, all subsequent computations of $\mathbf{C}_{kt}^{\mathrm{full}}$ for Step 1 can be performed recursively from the previous value using Eq.~\eqref{eq:smw}.

Further computational simplifications are possible by noting that the trace values required for computing Step 2 are also easily evaluated in a recursive fashion:
\begin{equation}
\begin{split}
\mathrm{Trace}\left[\mathbf{C}_{kt}^{\mathrm{full} \setminus \Xi_\ell}\right]& = \mathrm{Trace}\left[\mathbf{C}_{kt}^{\mathrm{full}}\right]+ \\
&\hspace{-.3in}\mathrm{Trace}\left[\mathbf{G}_{\ell t}^k\left(\mathbf{I}_C - \tilde{\mathbf{B}}_{\ell t}^k  \mathbf{G}_{\ell t}^k \right)^{-1} \left(\mathbf{G}_{\ell t}^k\right)^H\right].
\end{split}
\end{equation}

\section{Empirical Evaluations and Comparisons}\label{sec:results}
In the following subsections, we illustrate the performance and flexibility of OEDIPUS relative to traditional sampling design methods in several settings.

\subsection{2D T2-weighted Brain Imaging}\label{sec:t2}

In a first set of evaluations, we designed sampling patterns for the context of 2D T2-weighted brain imaging, based on an acquisition protocol that is routinely used at our neuroscience imaging center.    Data was acquired on the same scanner from seven different healthy volunteers on different days.\mycomment{AE.4}

\begin{figure}[tp]
\centering 
{\includegraphics[width=.8in]{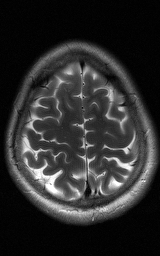}}
{\includegraphics[width=.8in]{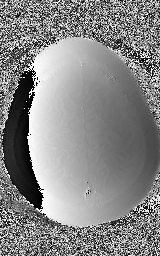}}
{\includegraphics[width=.8in]{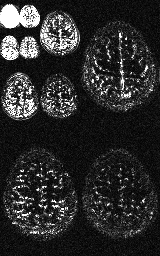}}\\
\subfigure[Magnitude]{\includegraphics[width=.8in]{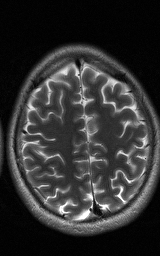}}
\subfigure[Phase]{\includegraphics[width=.8in]{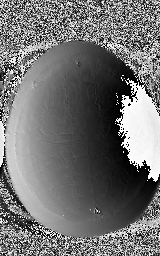}}
\subfigure[Wavelet]{\includegraphics[width=.8in]{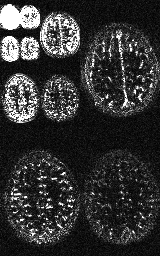}}
\caption{Gold standard reference images for the 2D T2-weighted brain imaging experiment.  The top row shows data from the first subject, while the bottom row shows data from the second subject.}
\label{fig:goldt2}
\end{figure}

All acquisition parameters were identical for all subjects. Imaging data was acquired with a TSE pulse sequence on a 3T scanner using a 12-channel receiver coil.   We used a fully-sampled 256$\times$160 Nyquist-rate Fourier acquisition matrix, corresponding to a 210mm$\times$131mm FOV, with a 3.5mm slice thickness. Other parameters include TE=88ms and TR=10sec.  This data was whitened using standard methods \cite{pruessmann2001}, and receiver coil sensitivity maps were obtained using a smoothness-regularized estimation formulation \cite{allison2013}.  Gold standard reference images were obtained by applying SENSE reconstruction \cite{pruessmann1999,pruessmann2001} to the fully sampled data.  These gold standard images for two subjects are shown in Fig.~\ref{fig:goldt2}(a,b).

Sampling patterns were designed using OEDIPUS for two different imaging scenarios: single-channel imaging (a simple Fourier transform with no sensitivity encoding) and 12-channel imaging (including coil sensitivity maps). In both cases, the design was performed with respect to a single slice for simplicity, { although\mycomment{AE.1, R2.1} it would have been straightforward to use volumetric training data (or, to reduce computational complexity, perhaps a small number of selected slices representing the range of expected variations across the volume of interest) for cases where broader volume coverage may be desired}.\mycomment{AE.4}  We denote the OEDIPUS sampling patterns designed for the single-channel  and multi-channel cases as SCO and MCO, respectively.  Since this is a 2D imaging scenario, OEDIPUS was constrained so that all samples from the same readout line must be sampled simultaneously, meaning that undersampling was only performed in 1D along the phase encoding dimension.   

 For illustration (and without loss of generality),\mycomment{AE.1} we show OEDIPUS sampling patterns that were designed for a single examplar (i.e., $K=1$), and in the multi-channel case, for a single set of sensitivity maps (i.e., $T=1$).  Specifically, OEDIPUS was performed based on the transform-domain sparsity pattern and sensitivity maps estimated from the first subject.  Sparse approximation of this dataset was obtained by keeping the top 15\% largest-magnitude transform coefficients (i.e., $S=0.15 N$).  The sparsifying transform $\boldsymbol{\Psi}$ was chosen to be the 3-level wavelet decomposition using a Daubechies-4 wavelet filter, whose transform coefficients are shown for the first two subjects in Fig.~\ref{fig:goldt2}(c).    As can be seen, while the images and the exact distributions of wavelet coefficients are distinct for each of the two datasets, the wavelet-domain support patterns for both images still have strong qualitative similarities to one another.  This similarity is one of the key assumptions that enables OEDIPUS to design sampling patterns based on exemplar data instead of requiring subject-specific optimization.

MCO sampling patterns were designed for the acceleration factors of $R=2$, 3, and 4, and an SCO sampling pattern was designed for $R=2$.\mycomment{AE.4} We also attempted to design SCO sampling patterns for the acceleration factors of $R=3$ and 4, but did not obtain meaningful OEDIPUS results because the matrix inversions needed for computing the CRB became numerically unstable.  This instability occured because the matrices became nearly-singular, which is an indication that acceleration factors of 3 or 4 may be too aggressive for unbiased estimation in this particular scenario. Specifically, near-singularity implies that unbiased estimators would have very large variance for certain parameters of interest (i.e., a singular matrix is associated with infinite variance \cite{stoica2001}).  This is consistent with our empirical results (not shown due to space constraints) for which $R=3$ or 4 yields unacceptably poor results in this scenario for every case we attempted.

For comparison against SCO and MCO, we also considered two classical undersampling patterns:  uniform undersampling and random undersampling.  Uniform undersampling is a deterministic sampling pattern that is frequently used in parallel imaging \cite{pruessmann1999}, and obtains phase encoding lines that are uniformly spaced on a regular lattice in k-space.  Random sampling is often advocated for sparsity-constrained reconstruction \cite{lustig2007,haldar2010a,wang2010,knoll2011a,zijlstra2016}.  In this case, we used a common approach in which the central 16 lines of k-space were fully sampled, and the remaining portions of k-space were acquired with uniform Poisson disc random sampling \cite{nayak1998}.    It is worth mentioning that both the uniform and random sampling schemes considered in this work are expected to be better-suited to multi-channel scenarios than they are to single-channel cases.  In particular, uniform undersampling is not usually advocated in single-channel settings due to the well-known coherent aliasing associated with lattice sampling, and single-channel random undersampling usually employs a more gradual variation in sampling density rather than the abrupt binary change in sampling density that we've implemented.  Nevertheless, due to the large number of subjects and scenarios we consider in this paper and the effort associated with manually tuning more nuanced sampling density variations  by trial-and-error, we have opted for the sake of simplicity to utilize the same type of simple random sampling design for both single- and multi-channel cases.  To avoid the possibility of getting a particularly bad random design, we created ten independently generated realizations of Poisson disc sampling, and report the best quantitative results.

\begin{figure}[tp]
\centering 
\begin{tabular}{c>{\centering\arraybackslash}m{0.72in} >{\centering\arraybackslash}m{0.72in} >{\centering\arraybackslash}m{0.72in} >{\centering\arraybackslash}m{0.72in}}
& Uniform & Random & SCO & MCO\\
$R=2$ & \fcolorbox{red}{red}{\includegraphics[width=0.7in]{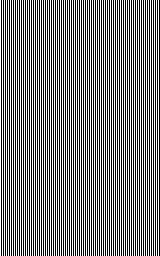}} & \fcolorbox{OliveGreen}{OliveGreen}{\includegraphics[width=0.7in]{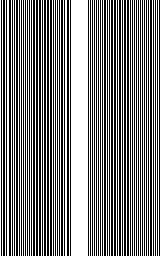}} & \fcolorbox{blue}{blue}{\includegraphics[width=0.7in]{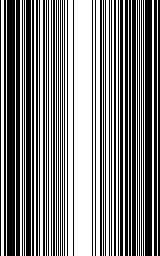}} & \fcolorbox{violet}{violet}{\includegraphics[width=0.7in]{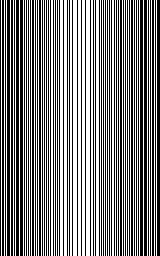}}  \\
$R=3$ & \fcolorbox{red}{red}{\includegraphics[width=0.7in]{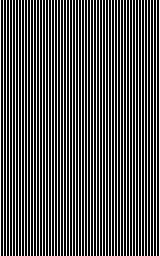}} & \fcolorbox{OliveGreen}{OliveGreen}{\includegraphics[width=0.7in]{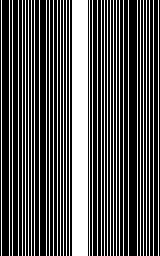}} &  & \fcolorbox{violet}{violet}{\includegraphics[width=0.7in]{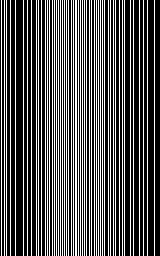}} \\
$R=4$ & \fcolorbox{red}{red}{\includegraphics[width=0.7in]{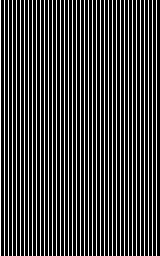}} & \fcolorbox{OliveGreen}{OliveGreen}{\includegraphics[width=0.7in]{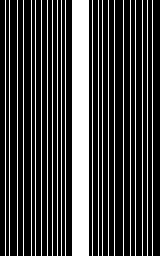}} &  & \fcolorbox{violet}{violet}{\includegraphics[width=0.7in]{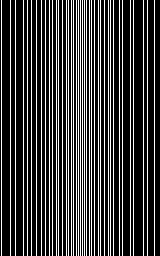}} 
\end{tabular}
\caption{Sampling patterns for the 2D T2-weighted brain imaging experiment.}
\label{fig:sampt2}
\end{figure}

The sampling patterns we obtained are shown in Fig.~\ref{fig:sampt2}, and demonstrate some interesting features.  In particular, we observe that both SCO and MCO generate variable density sampling patterns, which is consistent with the popularity of variable density sampling approaches in the literature \cite{lustig2007,haldar2010a,wang2010,knoll2011a,zijlstra2016}.  However, unlike standard randomization-based variable density approaches, we observe that OEDIPUS automatically tailors the sampling pattern to the specific imaging context, without the need for manual interaction.  This can be observed from the fact that OEDIPUS automatically produces different sampling density variations for single-channel and multi-channel contexts, with SCO sampling the center of k-space more densely than MCO.  

To evaluate these different sampling patterns in a practical  reconstruction context, we also performed sparsity-constrained reconstructions of retrospectively-undersampled single-channel and multi-channel datasets derived from the gold standard reference images.      We might expect to see especially good empirical reconstruction results for OEDIPUS for the first subject, since the SCO and MCO sampling patterns were designed based on the characteristics of that data.  In this case, we are testing on the same data that we trained on, and our results may be overly optimistic due to the common problem of overfitting.  On the other hand, the empirical results from the other subjects may be more interesting, { since these  will demonstrate the capability to design sampling patterns that generalize beyond the training data.}\mycomment{AE.1}

We performed two different types of sparsity-constrained reconstruction.  In the first case, we applied standard $\ell_1$-regularized reconstruction \cite{lustig2007} according to
\begin{equation}
\hat{\mathbf{f}} = \arg\min_{\mathbf{f} \in \mathbb{C}^N} \|\mathbf{A}\mathbf{f} - \mathbf{d}\|_2^2 + \lambda \|\boldsymbol{\Psi}\mathbf{f}\|_1.\label{eq:opt}
\end{equation}
This formulation will encourage sparsity in the 3-level Daubechies-4 wavelet decomposition of the image.  The regularization parameter $\lambda$ was set to a small number (i.e., $\lambda=0.01$) to promote data consistency.\mycomment{AE.4}  We might expect OEDIPUS to perform particularly well in this case, because the sparsifying transform used to design SCO and MCO sampling patterns is matched to the sparsifying transform being used for reconstruction.

While MR images are known to posses wavelet sparsity, it has also been observed that wavelet-based sparse regularization often results in undesirable high-frequency oscillation artifacts.  As a result, many authors (e.g., \cite{lustig2007,haldar2010a}) have advocated the use of total variation (TV) regularization \cite{rudin1992} to either augment or replace wavelet regularization.  TV regularization is designed to impose sparsity constraints on the finite differences of a reconstructed image.  Unfortunately, our OEDIPUS formulation is not easy to apply to TV regularization, because TV is not associated with a left-invertible transform.  However,  some authors have postulated that CRB-based experiment design may still be beneficial even in cases where the model used to compute the CRB is different from the model used for image reconstruction \cite{marseille1994}, as long as both models capture the behavior of similar image features of interest.  To test this, we also performed TV-regularized reconstruction.  Specifically, we solved the same optimization problem as in Eq.~\eqref{eq:opt}, but replacing the $\ell_1$-norm with a TV penalty while leaving the regularization parameter $\lambda$ the same.

In both cases, optimization was performed using a multiplicative half-quadratic algorithm \cite{nikolova2005, haldar2011c} (also known as ``Iteratively Reweighted Least Squares'' or ``Lagged Diffusivity'').  Reconstruction quality was measured quantitatively using the normalized root-mean-squared error (NRMSE), defined as $\text{NRMSE} = \|\hat{\mathbf{f}}-\mathbf{f}^*\|_2/\|\mathbf{f}^*\|_2,$ where $\mathbf{f}^*$ is the gold standard.

The empirical NRMSE results for the reconstruction of single-channel data for the first two subjects are shown in Table~\ref{tab:sct2}  and for the remaining subjects in Supplementary Table~SI, with illustrative reconstructed images shown in Supplementary Fig.~S1,\footnote{This paper has supplementary downloadable material provided by the authors, available at http://ieeexplore.ieee.org in the supplementary files/multimedia tab.} while NRMSE results for the reconstruction of multi-channel data are shown for the first two subjects in Table~\ref{tab:mct2}  and for the remaining subjects in Supplementary Table~SII,  with illustrative reconstructed images shown in Supplementary Fig.~S2. 

\begin{table}[tp]
\renewcommand{\arraystretch}{1.3}
\caption{Table of NRMSE values for reconstructed Single-Channel 2D T2-weighted brain data. Results are shown for Wavelet (Wav) and Total Variation (TV) reconstruction approaches, and for uniform (Uni), random (Rand), and OEDIPUS sampling patterns.}
\centering
\begin{tabular}{c|c|cccc|cccc|}
\cline{3-10}
\multicolumn{2}{c|}{} & \multicolumn{4}{c|}{\bfseries First Subject (Training)} & \multicolumn{4}{c|}{\bfseries  Second Subject}\\
\cline{3-10}
 \multicolumn{2}{c|}{} & \bfseries Uni &\bfseries Rand & \bfseries SCO & \bfseries MCO & \bfseries Uni & \bfseries Rand & \bfseries SCO & \bfseries MCO \\
\hline
\multirow{2}{*}{$R=2$} & \bfseries Wav & 0.792 & 0.178 & \bfseries 0.126 & 0.310 & 0.855 & 0.204 & \bfseries  0.143 & 0.339 \\
& \bfseries TV &  0.801 & 0.114 & {\bf 0.095} & 0.185 & 0.801 & 0.155 & {\bf 0.114} & 0.305 \\\hline 
\end{tabular}\label{tab:sct2}
\end{table}

\begin{table}[tp]
\renewcommand{\arraystretch}{1.3}
\caption{Table of NRMSE values for reconstructed Multi-Channel 2D T2-weighted brain data. Results are shown for Wavelet (Wav) and Total Variation (TV) reconstruction approaches, and for uniform (Uni), random (Rand), and OEDIPUS sampling patterns.}
\centering
\begin{tabular}{c|c|cccc|cccc|}
\cline{3-10}
\multicolumn{2}{c|}{} & \multicolumn{4}{c|}{\bfseries First Subject (Training)} & \multicolumn{4}{c|}{\bfseries  Second Subject}\\
\cline{3-10}
 \multicolumn{2}{c|}{} & \bfseries Uni &\bfseries Rand & \bfseries SCO & \bfseries MCO & \bfseries Uni & \bfseries Rand & \bfseries SCO & \bfseries MCO \\
\hline
\multirow{2}{*}{$R=2$} & \bfseries Wav & \bfseries 0.046 & 0.051 & 0.056 & 0.049 & \bfseries 0.053 & 0.061 & 0.065 &  0.056 \\
& \bfseries TV & \bfseries 0.043 &  0.046 & 0.052 &  0.046 &  \bfseries 0.050 & 0.055 & 0.060 & 0.053 \\\hline
\multirow{2}{*}{$R=3$} & \bfseries Wav & 0.087 & 0.090 & & \bfseries 0.072 & 0.105 & 0.110 & & \bfseries 0.084 \\
& \bfseries TV &  0.071 & 0.076 & & \bfseries 0.064 & 0.085 & 0.092 & &  \bfseries 0.075\\ \hline
\multirow{2}{*}{$R=4$} & \bfseries Wav & 0.160 & 0.149 & & \bfseries 0.096 & 0.194 & 0.176 & & \bfseries 0.113 \\
& \bfseries TV &  0.126 & 0.117 & & \bfseries 0.082 & 0.145 & 0.144 & &  \bfseries 0.097 \\
\hline
\end{tabular}
\label{tab:mct2}
\end{table}

For single-channel data, the SCO sampling pattern worked the best for both wavelet- and TV-based reconstruction.    The differences in NRMSE between SCO and the other approaches were often substantial, with random sampling being the next best approach.  Uniform undersampling had very poor performance in this case, although this is unsuprising because uniform undersampling is known to be a poor choice for sparsity-constrained reconstruction of single-channel data \cite{lustig2007}.  

In the multi-channel case, the MCO sampling worked the best for both wavelet- and TV-based reconstruction at high acceleration factors (i.e., $R=3$ and 4).  We observed that uniform random sampling worked the best with $R=2$, although MCO follows close behind and all four sampling schemes work similarly well at this low acceleration rate. The good performance of uniform undersampling for multi-channel data at low-acceleration rates occurs despite the fact that it had a slightly worse CRB than MCO, although is consistent with theoretical g-factor arguments from the parallel imaging literature \cite{pruessmann1999,pruessmann2001}.

We also observe that the SCO pattern seems to work better with single-channel data and MCO works better with multi-channel data, as should be expected.  In addition, our results show that the OEDIPUS sampling patterns designed for one subject generally yielded good performance when applied to similar data from other subjects.

\subsection{3D T1-weighted Brain Imaging}

In a second set of evaluations, we designed sampling patterns for the context of 3D T1-weighted brain imaging, also based on an acquisition protocol that is routinely used at our neuroscience imaging center.  Data was acquired on the same scanner from one healthy volunteer and one subject with a chronic stroke lesion.  The healthy volunteer was scanned more than 13 months after the stroke subject.

Most acquisition parameters were identical for both subjects.  Imaging data was acquired with an MPRAGE pulse sequence on a 3T scanner using a 12-channel receiver coil.  Along the two phase-encoding dimensions, we used a fully-sampled 156$\times$192 Fourier acquisition matrix, corresponding to a 208mm$\times$256mm FOV.  Because this FOV is much larger than necessary along the left-right dimension, all image reconstructions and experiment designs were performed with respect to a 156 $\times$ 126 voxel grid corresponding to a smaller 208 mm $\times$ 168 mm FOV.  The third (readout) dimension was reconstructed with 1mm resolution, and we extracted one slice along this dimension to be used in our sampling investigation.  Other parameters include TI=800ms, TE=3.09ms, TR=2530ms, and flip angle = 10$^\circ$.

\begin{figure}[tp]
\centering 
{\includegraphics[width=.8in]{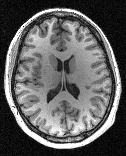}}
{\includegraphics[width=.8in]{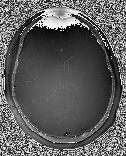}}
{\includegraphics[width=1.219047619in]{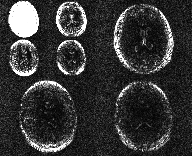}}\\
\subfigure[Magnitude]{\includegraphics[width=.8in]{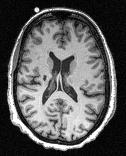}}
\subfigure[Phase]{\includegraphics[width=.8in]{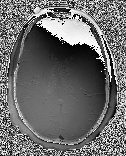}}
\subfigure[Wavelet]{\includegraphics[width=1.219047619in]{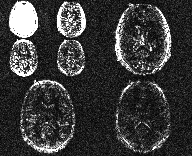}}
\caption{Gold standard reference images for the 3D T1-weighted brain imaging experiment.  The top row shows data from the healthy subject, while the bottom row shows data from the stroke subject.  The slice we've chosen for the stroke subject includes both a stroke lesion and a fiducial marker, both of which are absent in the image for the healthy subject.  The fiducial marker appears as a white circle  on the top left side of the image, while the lesion appears as a dark spot in the white matter near the middle of the brain on the left side of the image.}\label{fig:goldt1}
\end{figure}

While both datasets were acquired using the same 12-channel array coil, the data for the stroke subject had been hardware-compressed by the scanner down to 4 virtual channels, while we had access to the original 12-channels of data for the healthy subject.  To reduce the differences in sensitivity encoding information content between the two datasets, we used standard software coil-compression methods  \cite{buehrer2007,huang2008} to reduce the data for the healthy subject to 5 virtual channels.  From the coil-compressed data, we obtained receiver coil sensitivity maps and gold standard reference images following the same procedures as described in the previous section.  These gold standard images are shown in Fig.~\ref{fig:goldt1}.

Similar to the previous case, we used OEDIPUS to design sampling patterns for both single-channel imaging and multi-channel imaging.  Since this is a 3D imaging scenario, we gave OEDIPUS the freedom to undersample in 2D along both phase encoding dimensions.  { For illustration (and without loss of generality)},\mycomment{AE.1} OEDIPUS sampling patterns were designed using the same approach described in the previous section (i.e., $K=1$, $T=1$, $S=0.15N$, and using the 3-level Daubechies-4 wavelet decomposition) based on the  single-slice from the healthy subject.  The wavelet coefficients for both reference images are shown in Fig.~\ref{fig:goldt1}(c).  As before, we observe strong qualitative similarities in the wavelet-domain support patterns for both images, despite the presence of a lesion in one image but not in the other.

MCO and SCO sampling patterns were both designed for several acceleration factors between $R=2$ and 8.  Compared to the previous case, the use of 2D undersampling instead of 1D undersampling enables higher acceleration factors for SCO, although high ill-posedness still prevented us from computing meaningful SCO results for $R=7$ and 8.

Similar to the previous case, we compared SCO and MCO sampling designs against uniform and random sampling.  For uniform undersampling, we used CAIPI-type uniform lattice sampling designs that are known to have good parallel imaging characteristics \cite{breuer2006}.  For random sampling, we used a standard approach in which the central $16\times 16$ region of k-space was fully sampled, and the remaining portions of k-space were acquired with uniform Poisson disc random sampling.  As before, we created ten independently generated realizations of Poisson disc sampling, and report the best results.

\begin{figure}[tp]
\begin{tabular}{c>{\centering\arraybackslash}m{0.72in} >{\centering\arraybackslash}m{0.72in} >{\centering\arraybackslash}m{0.72in} >{\centering\arraybackslash}m{0.72in}}
& Uniform & Random & SCO & MCO\\
$R=2$ & 
\fcolorbox{red}{red}{\includegraphics[width=0.7in]{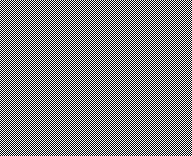}} & 
\fcolorbox{OliveGreen}{OliveGreen}{\includegraphics[width=0.7in]{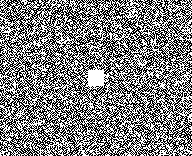}} & 
\fcolorbox{blue}{blue}{\includegraphics[width=0.7in]{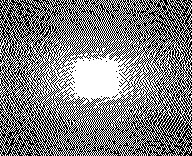}} & 
\fcolorbox{violet}{violet}{\includegraphics[width=0.7in]{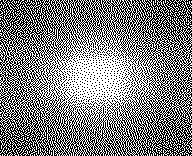}}  \\
$R=3$ & 
\fcolorbox{red}{red}{\includegraphics[width=0.7in]{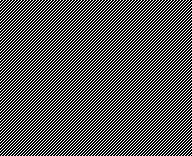}} & 
\fcolorbox{OliveGreen}{OliveGreen}{\includegraphics[width=0.7in]{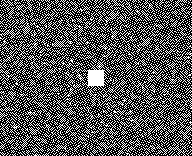}} &  
\fcolorbox{blue}{blue}{\includegraphics[width=0.7in]{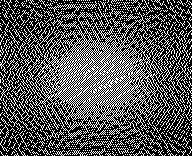}}& 
\fcolorbox{violet}{violet}{\includegraphics[width=0.7in]{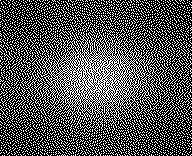}} \\
$R=4$ & 
\fcolorbox{red}{red}{\includegraphics[width=0.7in]{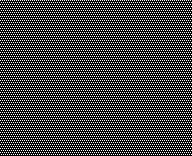}} & 
\fcolorbox{OliveGreen}{OliveGreen}{\includegraphics[width=0.7in]{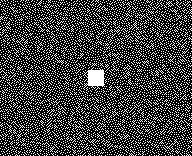}} &  
\fcolorbox{blue}{blue}{\includegraphics[width=0.7in]{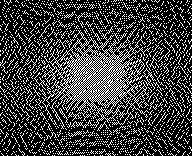}}& 
\fcolorbox{violet}{violet}{\includegraphics[width=0.7in]{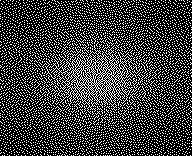}}  \\
$R=5$ & 
\fcolorbox{red}{red}{\includegraphics[width=0.7in]{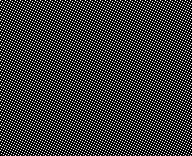}} & 
\fcolorbox{OliveGreen}{OliveGreen}{\includegraphics[width=0.7in]{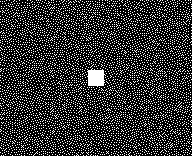}} &  
\fcolorbox{blue}{blue}{\includegraphics[width=0.7in]{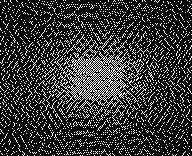}}& 
\fcolorbox{violet}{violet}{\includegraphics[width=0.7in]{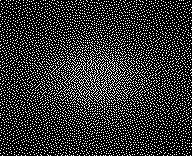}} \\
$R=6$ & 
\fcolorbox{red}{red}{\includegraphics[width=0.7in]{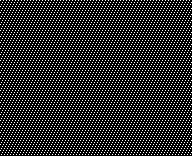}} & 
\fcolorbox{OliveGreen}{OliveGreen}{\includegraphics[width=0.7in]{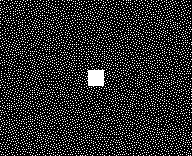}} &  
\fcolorbox{blue}{blue}{\includegraphics[width=0.7in]{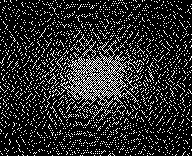}}& 
\fcolorbox{violet}{violet}{\includegraphics[width=0.7in]{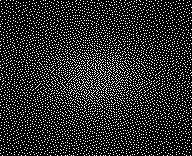}} \\
$R=7$ & 
\fcolorbox{red}{red}{\includegraphics[width=0.7in]{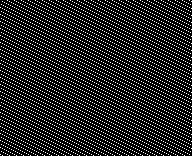}} & 
\fcolorbox{OliveGreen}{OliveGreen}{\includegraphics[width=0.7in]{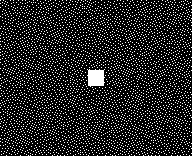}} &  
&\fcolorbox{violet}{violet}{\includegraphics[width=0.7in]{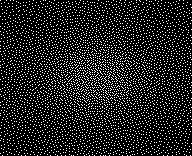}} \\
$R=8$ & 
\fcolorbox{red}{red}{\includegraphics[width=0.7in]{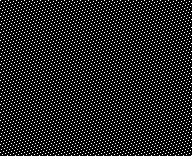}} & 
\fcolorbox{OliveGreen}{OliveGreen}{\includegraphics[width=0.7in]{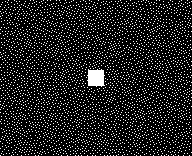}} &  
&
\fcolorbox{violet}{violet}{\includegraphics[width=0.7in]{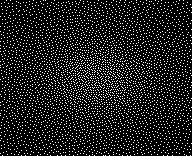}} 
\end{tabular}
\caption{Sampling patterns for the 3D T1-weighted brain imaging experiment.}\label{fig:sampt1}
\end{figure}

The sampling patterns we obtained are shown in Fig.~\ref{fig:sampt1}.  Similar to the previous case, we observe that both SCO and MCO generate variable density sampling patterns, but that SCO and MCO have very different sampling density characteristics from one another.  In particular, SCO frequently samples the center of k-space more densely than MCO does, while MCO samples peripheral k-space more densely than SCO.  Interestingly, we observe that SCO with $R=2$ performs full sampling of central k-space.  Because the  k-space acquisition grid was designed for a larger FOV while reconstruction is performed over a smaller FOV, this corresponds to oversampling the center of k-space  (i.e., at a rate higher than the conventional Nyquist rate).  On the other hand, at higher acceleration factors, we observe that both the SCO and MCO approaches sample the center of k-space less densely than conventional variable-density random sampling schemes (where the center of k-space is frequently sampled at the Nyquist rate) that are popular for sparsity-constrained MRI.  In addition, the SCO pattern visually appears to have more structure than is typical in  random designs.  This illustrates the fact that not only is OEDIPUS adaptive to the imaging context, but it can also explore a broader range of sampling patterns than traditional sampling design techniques.

Similar to before,  single-slice images were reconstructed from retrospectively undersampled data in both single-channel and multi-channel reconstruction contexts using either $\ell_1$-regularization of the wavelet transform coefficients or TV regularization.  The empirical NRMSE results for the reconstruction of single-channel data are shown in Table~\ref{tab:sct1} with illustrative reconstructed images shown in Supplementary Fig.~S3, while NRMSE results for the reconstruction of multi-channel data are shown in Table~\ref{tab:mct1} with illustrative reconstructed images shown in Supplementary Fig.~S4.\mycomment{AE.4}

\begin{table}[tp]
\renewcommand{\arraystretch}{1.3}
\caption{Table of NRMSE values for reconstructed Single-Channel 3D T1-weighted brain data. Results are shown for Wavelet (Wav) and Total Variation (TV) reconstruction approaches, and for uniform (Uni), random (Rand), and OEDIPUS  sampling patterns.}
\centering
\begin{tabular}{c|c|cccc|cccc|}
\cline{3-10}
\multicolumn{2}{c|}{} &  \multicolumn{4}{c|}{\bfseries Healthy Subject (Training)} & \multicolumn{4}{c|}{\bfseries  Stroke Subject}\\
\cline{3-10}
\cline{3-10}
 \multicolumn{2}{c|}{} & \bfseries Uni &\bfseries Rand & \bfseries SCO & \bfseries MCO & \bfseries Uni & \bfseries Rand & \bfseries SCO & \bfseries MCO \\
\hline
\multirow{2}{*}{$R=2$} & \bfseries Wav & 0.470 & 0.153 & \bfseries 0.117 & 0.118 & 0.413 & 0.122 & \bfseries 0.085 & 0.086 \\
& \bfseries  TV & 0.325 & 0.141 & \bfseries 0.104 & 0.105 & 0.238 & 0.114 & \bfseries 0.075 & 0.076 \\ \hline
\multirow{2}{*}{$R=3$} & \bfseries Wav & 0.601 & 0.249 & 0.199 & \bfseries 0.166 & 0.616 & 0.217 & 0.136 & \bfseries 0.135 \\
& \bfseries  TV & 0.512 & 0.203 & 0.145 & \bfseries 0.141 & 0.427 & 0.178 & \bfseries 0.110 & 0.112 \\ \hline
\multirow{2}{*}{$R=4$} & \bfseries Wav & 0.831 & 0.324 & \bfseries 0.270 & 0.650 & 0.856 & 0.295 & \bfseries 0.210 & 0.613 \\
& \bfseries  TV & 0.748 & 0.246 & \bfseries 0.182 & 0.289 & 0.867 & 0.233 & \bfseries 0.140 & 0.253\\ \hline 
\multirow{2}{*}{$R=5$} & \bfseries Wav &  0.815 & 0.389 & \bfseries 0.324 & 0.735 & 0.913 & 0.354 & \bfseries 0.273 & 0.699 \\
& \bfseries  TV & 0.732 & 0.288 & \bfseries 0.213 & 0.376 & 0.869 & 0.272 & \bfseries 0.177 & 0.371\\ \hline  
\multirow{2}{*}{$R=6$} & \bfseries Wav & 0.820 & 0.439 & \bfseries 0.402 &  0.767 & 0.896 & 0.383 & \bfseries 0.330 & 0.763 \\
& \bfseries  TV & 0.787 & 0.317 & \bfseries 0.264 & 0.535 & 0.889 & 0.297 & \bfseries 0.248 & 0.600\\ \hline 
\end{tabular}
\label{tab:sct1}
\end{table}

Although the use of 2D undersampling admits the use of higher acceleration factors than were achieved in the previous case, our observations are still largely consistent with the observations from Section~\ref{sec:t2}.  In particular, we observe that SCO generally yields the best performance in the single-channel case for both wavelet and TV-regularization, and for both the healthy subject (which OEDIPUS was trained for) and the stroke subject (which was not used for training).  Similarly, MCO generally yields the best performance in the multi-channel case in all of these different scenarios.  There are some exceptions to these general rules, e.g., MCO slightly outperforms SCO in many of the cases with $R=3$ and single-channel data, while MCO is outperformed by uniform (CAIPI) sampling in the multi-channel case with lower acceleration factors (despite having slightly better CRB values). This result is not necessarily surprising, since we have no guarantees that the OEDIPUS sampling design will yield optimal results for any given dataset.  However, it is noteworthy that in each of these cases where the most relevant OEDIPUS strategy was outperformed by an alternative approach, the OEDIPUS design is not far behind the leader.  In addition, at higher and higher acceleration factors, we observe that the most relevant OEDIPUS strategy consistently yields the best performance (by wider  margins as acceleration increases).

\begin{table}[tp]
\renewcommand{\arraystretch}{1.3}
\caption{Table of NRMSE values for reconstructed Multi-Channel 3D T1-weighted brain data. Results are shown for Wavelet (Wav) and Total Variation (TV) reconstruction approaches, and for uniform (Uni), random (Rand), and OEDIPUS  sampling patterns.}
\centering
\begin{tabular}{c|c|cccc|cccc|}
\cline{3-10}
\multicolumn{2}{c|}{} &  \multicolumn{4}{c|}{\bfseries Healthy Subject (Training)} & \multicolumn{4}{c|}{\bfseries  Stroke Subject}\\
\cline{3-10}
 \multicolumn{2}{c|}{} &  \bfseries Uni &\bfseries Rand & \bfseries SCO & \bfseries MCO & \bfseries Uni & \bfseries Rand & \bfseries SCO & \bfseries MCO \\
\hline
\multirow{2}{*}{$R=2$} & \bfseries Wav & \bfseries 0.070 & 0.074 & 0.074 & 0.073 & \bfseries 0.053 & 0.056 & 0.055 & 0.054 \\
& \bfseries  TV & \bfseries 0.069 & 0.071 & 0.071 & \bfseries 0.069 & \bfseries 0.052 & 0.054 & \bfseries 0.052 & \bfseries 0.052 \\
\hline
\multirow{2}{*}{$R=3$} & \bfseries Wav & \bfseries 0.092 & 0.095 & 0.103 & 0.096 & \bfseries 0.070 & 0.072 & 0.074 & 0.071 \\
& \bfseries  TV & \bfseries 0.088 & 0.089 & 0.092 & 0.089 & \bfseries 0.067 & 0.068 & 0.069 & \bfseries 0.067 \\
\hline
\multirow{2}{*}{$R=4$} & \bfseries Wav & \bfseries 0.112 &  0.114 & 0.126 &  0.114 & 0.087 & 0.089 & 0.090 & \bfseries 0.086\\
& \bfseries  TV & \bfseries 0.103 & 0.105 & 0.109 & \bfseries 0.103 & 0.080 & 0.082 & 0.082 & \bfseries 0.079 \\
\hline
\multirow{2}{*}{$R=5$} & \bfseries Wav & \bfseries 0.122 & 0.129 & 0.142 & 0.129 & 0.104 & 0.107 & 0.105 & \bfseries 0.098 \\
& \bfseries  TV & \bfseries 0.112 & 0.117 & 0.121 & 0.113 & \bfseries 0.089 & 0.095 & 0.093 & \bfseries 0.089 \\ \hline 
\multirow{2}{*}{$R=6$} & \bfseries Wav & 0.144 & 0.145 & 0.154 & \bfseries 0.141 &  0.161 & 0.126 & 0.117 & \bfseries 0.111\\
& \bfseries  TV & 0.128 & 0.129 & 0.130 & \bfseries 0.122 & 0.105 & 0.109 & 0.102 & \bfseries 0.098  \\ \hline 
\multirow{2}{*}{$R=7$} & \bfseries Wav & 0.181 & 0.161 & & \bfseries 0.153 & 0.316 & 0.151 & & \bfseries 0.122 \\
& \bfseries  TV & 0.143 & 0.140 & & \bfseries 0.130 &0.146 & 0.128 & & \bfseries 0.106 \\ \hline 
\multirow{2}{*}{$R=8$} & \bfseries Wav & 0.194 & 0.178 & & \bfseries 0.162 & 0.336 & 0.177 & & \bfseries 0.135 \\
& \bfseries  TV &  0.156 & 0.152 & & \bfseries 0.138 & 0.164 & 0.145 & & \bfseries 0.114\\ \hline 
\end{tabular}
\label{tab:mct1}
\end{table}

{ The \mycomment{R2.1, AE.3} sampling patterns we've shown were optimized for a single slice near the center of the brain, although it's reasonable to expect that the sampling patterns will still be reasonably good for the neighboring slices from the volume that have similar image features and coil sensitivity characteristics.  Reconstruction quality performance across the 3D volume is shown for different sampling patterns in Supplementary Figs.~S5 and S6, and match our expectations.  In particular, performance is very consistent for the slices that are in a similar region  to the slice used for training, although performance sometimes has larger deviations in farther slices that have more distinct characteristics (e.g., image regions that have very different support characteristics, or which have different textural features, etc.).  Note that even though our results generalize relatively well across the volume, we still do not recommend using single-slice training data if volumetric reconstruction is ultimately of interest, as this would likely lead to suboptimal results that are biased towards one anatomical region more than others.  We believe that it would be much preferred to optimize sampling across exemplar volumes, or at least across a few representative slices reflecting the variations across the volume.  The general principal of OEDIPUS is that the training data should be relatively well-matched to the anatomical regions that are of highest importance to the specific application for which the sampling pattern is  designed.}

\section{Discussion and Conclusions}\label{sec:disc}

This paper introduced the OEDIPUS framework to designing sampling patterns for sparsity-constrained MRI.  Unlike standard randomization based approaches (which have some major theoretical shortcomings \cite{haldar2010a}), our proposed approach is guided by estimation theory, is deterministic, and automatically tailors itself to the given imaging context at hand.  

OEDIPUS was evaluated empirically at many different acceleration factors in several very different imaging contexts.   While these empirical evaluations were limited in scope and should not be overinterpreted, our results were very consistent in each case, and suggest several potential conclusions.  First, results seem to confirm that the OEDIPUS-derived sampling patterns can be used to design good sampling patterns that outperform some of the classical sampling approaches.  While it may be possible to manually tune a variable density random sampling pattern that achieves similar or even better performance to the OEDIPUS patterns,  OEDIPUS has the advantage that the sampling density is deterministic and automatically tailored to the specific scenario of interest, without the need for the user to exhaustively explore a range of different random sampling distributions or realizations.    Second, sampling patterns designed based on data from one subject appear to generalize well to other subjects. This kind of generalization capability is important, since it would mean that it is not necessary to create unique sampling patterns for each subject, and that offline sampling pattern design may indeed be a viable strategy.   And finally, results suggest that OEDIPUS sampling schemes designed based on wavelet sparsity may generalize well to the case of TV reconstruction.  Given these results, we believe that OEDIPUS has the potential to  be a powerful tool for sampling design across a range of different applications.  

The OEDIPUS framework is based on the constrained CRB, which is an estimation theoretic measure for the amount of information provided by a given experiment design.  It should be noted that other metrics have also previously been used to measure the expected ``goodness'' of a sampling design in sparse MRI reconstruction, including mutual incoherence and restricted isometry constants \cite{lustig2007,haldar2010a}.  These other metrics have been previously considered because, when these mathematical constants are good enough, it is possible to guarantee the performance of sparsity-based reconstruction methods \cite{donoho2006,candes2008b}, and these guarantees will hold for arbitrary sparse images, regardless of their transform domain support characteristics.  In this sense, these are measures that can be used to characterize performance for the ``worst-case'' transform domain  support patterns. However, unlike OEDIPUS, these measures are agnostic to ``typical" realizations of the inverse problem.  Designing for the worst case is likely to sacrifice performance in typical cases.  In addition, the conditions under which these kinds of metrics could be used to provide strong performance guarantees are often not met in practical MRI applications \cite{haldar2010a}.

While the implementation described in this paper made use of the greedy SBS method for simplicity, we expect that even better OEDIPUS performance could be further enhanced using appropriate exchange algorithms \cite{broughton2010}.  With good initializations (e.g., initializing with CAIPI), these kinds of algorithms could also be used to decrease computation time and increase the chances of finding good local optima relative to SBS.

Since  OEDIPUS is intended for offline use (i.e., experiment designs only need to be computed once for a given imaging context and can then be reused ad infinitum), its associated computation time is perhaps less important than for other kinds of optimization procedures.  Nevertheless, it is worth noting that the computation time can be relatively mild or relatively burdensome depending on the imaging context.  At least for the SBS algorithm, the number of candidate sampling subsets $L$ has a very big impact on computation time.  For example, in the 2D T2-weighted illustration, we started with a relatively small number of candidates to consider ($L=160$) and optimization completed in a few hours on a standard desktop computer.  On the other hand, we started with a much larger number of candidates in the 3D T1-weighted illustration ($L=19,656$), and optimization in this case can require many days of computation.  The problem of large $L$ is not new in the experiment design literature, and various approaches have been suggested to reduce $L$.  For example, in a different context, Gao and Reeves \cite{gao2001} have suggested the use of simpler classes of sampling patterns (e.g., periodic nonuniform sampling) with fewer candidates to consider.  This improves computational complexity though sacrifices optimality.

While our results were shown in the context of static MRI imaging, our approach  generalizes in natural ways to other imaging contexts where sparsity constraints may be useful.  For example, this approach has the potential to be useful in MRI parameter mapping with sparsity constraints, for which CRBs have already been described \cite{zhao2014}, and  we have performed some initial investigations of OEDIPUS-like sampling \cite{kim2017b} in the context of high-dimensional diffusion-relaxation correlation spectroscopic imaging \cite{kim2017a,kim2018}.

Although OEDIPUS was formulated to address sparsity constraints, there are natural generalizations to other imaging constraints that are recently popular in the literature.  For example, CRBs have been derived for low-rank matrix recovery \cite{tang2011}, structured low-rank matrix recovery \cite{zachariah2012}, and low-rank tensor recovery \cite{liu2001}, and it would be straightforward in principle to adapt the OEDIPUS approach to design sampling patterns for popular recent MRI reconstruction methods that leverage such low-rank modeling assumptions \cite{haldar2010,goud2011,haldar2011,shin2014,haldar2013b,haldar2017a,trzasko2013b,he2016}.  This promises to be an interesting direction for further research.

{ The\mycomment{R2.2} OEDIPUS formulation we used in this work was designed to minimize the average variance, which is a parameter related to NRMSE.  Since NRMSE has some well-known limitations \cite{kim2018a}, the  exploration of alternative  criteria  appears worthwhile.\footnote{During the review of this paper and after public dissemination  of OEDIPUS \cite{haldar2017,haldar2018}, a preprint appeared that describes an alternative method for optimizing k-space sampling patterns based on exemplar data \cite{gozcu2018}.  This approach has many similarities to OEDIPUS but is able to address other error metrics besides NRMSE, at the cost of higher computational complexity.}\mycomment{AE.4}  We believe that focused application-specific testing is especially important, since the characteristics of nonlinear methods can be hard to generalize from one context to the next \cite{haldar2010a}, and because  important context-specific image features may not correlate with NRMSE.}

As a final note, we believe it is worth reiterating that, despite common misconceptions, sparsity-constrained MRI generally does not enjoy any meaningful theoretical performance guarantees \cite{haldar2010a}.  OEDIPUS is no exception, and while OEDIPUS was observed to work well in several practical scenarios, we make no claim that OEDIPUS will always be optimal.

\bibliographystyle{IEEEtran}
\bibliography{./bibliography}

\clearpage\newpage
{\color{white} .}
\includepdf[pages=-]{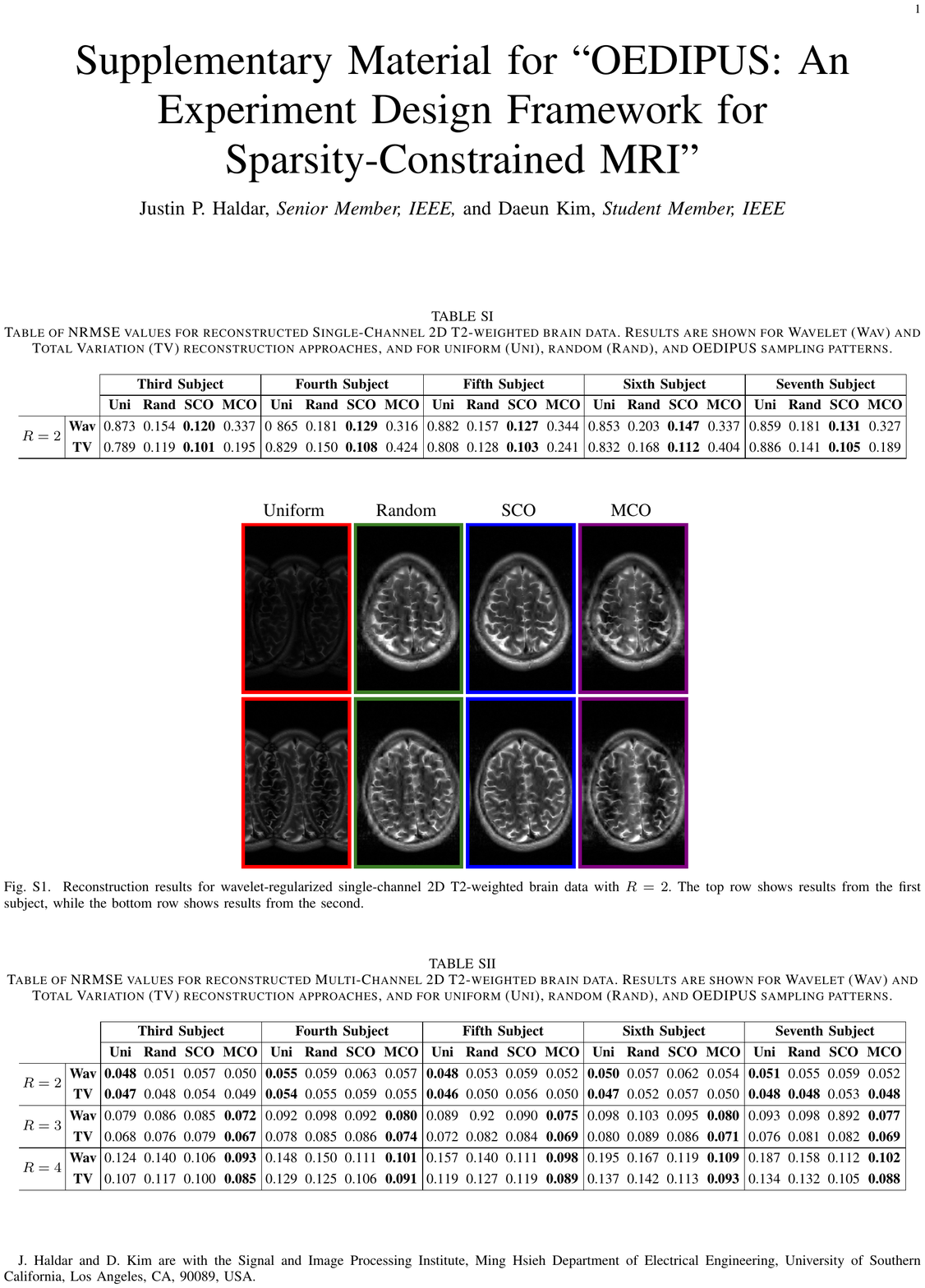}
\end{document}